\documentclass{emulateapj}
\usepackage{placeins}
\usepackage{graphicx}
\usepackage{amsmath}
\shorttitle{Particles Trapped at Gap Edges and Vortices: MHD cases} \shortauthors{Zhu et al.}

\newcommand{\del}{{\bf \nabla}}
\newcommand{\bmath}[1]{\mbox{\boldmath{$#1$}}}

\newcommand\msunyr{\rm M_{\odot}\,yr^{-1}}
\newcommand\be{\begin{equation}}
\newcommand\en{\end{equation}}

\newcommand\etal{{\rm et al}.\ }

\begin{document}

\title{DUST TRAPPING BY VORTICES 
IN TRANSITIONAL DISKS: EVIDENCE FOR NON-IDEAL MHD EFFECTS IN PROTOPLANETARY DISKS}

\author{Zhaohuan Zhu\altaffilmark{1,2}, AND  James M. Stone\altaffilmark{1}}

\altaffiltext{1}{Department of Astrophysical Sciences, 4 Ivy Lane, Peyton Hall,
Princeton University, Princeton, NJ 08544, USA}
\altaffiltext{2}{Hubble Fellow.}

\email{zhzhu@astro.princeton.edu }

\begin{abstract}
We study particle trapping at the edge of a gap opened by a planet
in a protoplanetary disk. In particular, we explore the effects of turbulence driven by the magnetorotational instability on particle trapping, using
global  three-dimensional magnetohydrodynamic (MHD) simulations including Lagrangian dust particles.
We study disks either in the ideal MHD limit or dominated by ambipolar diffusion (AD) that plays an essential role at the outer regions of a protoplanetary disk.
With ideal MHD, 
strong turbulence (the equivalent viscosity parameter  $\alpha\sim 10^{-2}$) in disks
prevents vortex formation at the 
 edge of the gap opened by a 9 $M_{J}$ planet, and most particles (except the particles that drift fastest) pile up at the outer gap edge almost axisymmetrically. 
When AD is considered, turbulence is significantly suppressed 
($\alpha\lesssim 10^{-3}$), and a large vortex
forms at the edge of the planet induced gap, which survives $\sim$ 1000 orbits. The vortex can efficiently trap dust particles that span 3 orders of magnitude in size
within 100 planetary orbits.  We have also
carried out two-dimensional hydrodynamical simulations using viscosity as an approximation to MHD turbulence. 
These hydrodynamical simulations can reproduce vortex generation at the gap edge as seen in MHD simulations. 
Finally, we use our simulation results to generate synthetic images for ALMA dust continuum observations 
on Oph IRS 48 and HD 142527, which show
good agreement with existing observations. Predictions for future ALMA cycle 2 observations have been made. 
We conclude that 
the asymmetry in ALMA observations can be explained by dust trapping vortices and the existence of vortices could be the evidence that the outer protoplanetary disks are dominated by ambipolar diffusion with  
$\alpha<10^{-3}$ at the disk midplane.
\end{abstract}


\section{Introduction}
{  Transitional disks are protoplanetary disks whose inner regions have undergone substantial clearing  
(see the review by Espaillat \etal 2014). 
Recent observations suggest that gas and dust are depleted at different levels at the inner regions of these disks.} Compared with
a full disk around a classical T Tauri star,
the dust surface density in the inner regions of transitional disks needs to be depleted by several orders of magnitude for producing 
the observed near-infrared deficit in their SEDs (Calvet \etal 2005). 
On the other hand, the gas
 surface density should be similar between the inner regions of transitional disks and those of
 full disks (Zhu \etal 2011), since transitional disks and full disks have similar
gas accretion rates (Najita \etal 2007; Espaillat \etal 2012). 
The presence of a significant amount of gas at the inner regions of transitional disks is confirmed by molecular line observations 
(Najita 2008; Ingleby \etal 2009; Salyk \etal 2009; France \etal 2012 ). Different distributions between dust and gas at the inner regions of transitional disks have also been 
supported by resolved disk observations at different wavelengths (e.g., near-infrared imaging  and submillimeter interferometry, 
Andrews \etal 2011; Dong \etal 2012;  Follette \etal 2013; Garufi \etal 2013), or with submillimeter interferometry using both  dust continuum and molecular lines (Casassus \etal 2013; van der Marel \etal 2013; Bruderer \etal 2014; Menu \etal 2014; Carmona \etal 2014).

{ Such different gas and dust distributions in transitional disks suggest that, besides gas dynamics, 
dust particle dynamics is also important during the process of disk clearing.}
There are two major
clearing processes proposed: gap opening by planets and disk photoevaporation 
(see the reviews by Kley \& Nelson 2012; Baruteau \etal 2013; Alexander \etal 2013). In
the gap opening by planets scenario, 
 gas can flow across the gap from the outer disk to the inner disk due to the  gravitational force from the planet,
explaining the high accretion rates of some transitional disks, while particles will be trapped at the outer
edge of the gap due to the aerodynamic gas drag (the ``dust filtration'' process, 
Rice \etal 2006; Zhu \etal 2012; Pinilla \etal 2012). Thus, the dust at the inner disk is more depleted than the gas,
which seems to be consistent with transitional disk observations.

{ Some transitional disks also show non-axisymmetric dust distributions 
at outer disks, revealed by recent ALMA observations} (Casassus \etal 2013;
van der Marel \etal 2013; Isella \etal 2013). In the extreme case of Oph IRS 48, there is a highly
asymmetric crescent-shaped dust structure between 45 and 80 AU from the star. 
The peak emission from this dust structure is at least 130  times stronger than the upper limit
of the opposite side of the disk. 

 {  To explain these non-axisymmetric dust structure in
transitional disks, particle trapping by vortices is proposed.} 
Anticyclonic vortices are long lived and can efficiently trap dust particles
(Barge \& Sommeria 1995; Adams \& Watkins 1995; Tanga \etal 1996; Bracco et al. 1999;
 Godon \& Livio 2000; Lyra \etal 2009ab; Johansen \etal 2004; Heng \& Kenyon 2010). Small scale 
 turbulence can affect particle trapping in vortices (Chavanis 2000) and may be important to explain recent ALMA observations (Lyra \& Lin 2013,
 P{\'e}rez et al. 2014).

{ Although vortices can be generated by various hydrodynamic instabilities in disks, in the scenario of gap opening
by planets, the gap edge is subject to the Rossby wave instability (RWI) which can naturally lead to vortex formation}
(Lovelace \etal 1999; Li \etal 2000; see the recent review by Lovelace \& Romanova 2013).
Spiral shocks excited by a planet push the
disk material away from the planet, leading to gap opening.
This process also piles up material at the gap edge, leading to a density bump which 
has a vortensity minimum and is subject to the RWI (Koller \etal 2003; Li \etal 2005; Li \etal 2009; Yu \etal 2010; Lin \& Papaloizou 2010). 
Recently, three-dimensional (3-D) simulations (Meheut \etal 2010; Lin 2012a, 2012b) with 
different equations of state (Lin 2013) have been carried out to study vortex formation by the RWI.  
Particle trapping in 3-D vortices has also been studied using 3-D hydrodynamical global simulations including dust dynamics  
 (Meheut \etal 2012, Zhu \etal 2014a), which suggests that particles at certain sizes could have  a factor of more than 100 increase in the dust surface density within the vortex.

{ However, all previous studies on particle trapping at the gap edge or vortices 
use inviscid hydrodynamic models or
viscous hydrodynamic models as an approximation to disk turbulence. In this work, we carry out direct MHD turbulent simulations to study
particle trapping at the gap edge and vortices with a planet in a turbulent disk. Our simulations have two new aspects compared with
previous work.}

{ First, we have considered dust particle dynamics in 
global turbulent gas disks.}
Dust particle distribution sensitively 
depends on gas distribution due to the gas drag. For example, an extremely shallow gap (e.g., $10\%$ perturbation to the disk surface density)
opened by a low mass planet (e.g., 10 $M_{\oplus}$) can trap particles at the gap edge efficiently (Zhu \etal 2014a).
In MHD simulations, turbulence driven by the magnetorotational instability (MRI, Balbus 
\& Hawley 1991) can also lead to significant particle concentration (Johansen \etal 2011).
Although previous MHD simulations, which study gap opening by
planets without including dust particles, have not reported strong gap edge vortices (Winters et al. 2003; Nelson \& 
Papaloizou 2003; Papaloizou et al. 2004; Uribe et al. 2011; Zhu \etal 2013; Gressel \etal 2013), it is unclear
if a weak vortex exists at the gap edge and be capable of trapping dust particles.

{ Second, we have considered the non-ideal MHD effect---ambipolar diffusion---in our global MHD simulations.} 
Transitional disks have wide gaps at 10s of {\rm AU}, where
Ambipolar diffusion (AD, Blaes \& Balbus 1994; Mac Low \etal 1995; Hawley \& Stone 1998; Kunz \& Balbus 2004) should be the dominant non-ideal MHD effect (Bai 2011a, 2011b; Perez-Becker \& Chiang 2011a, 2011b; Armitage \etal 2011; Turner \etal 2014). AD
can significantly suppress turbulence led by the MRI (Brandenburg et al. 1995; Bai \& Stone 2011).
By using viscosity to approximate MRI turbulence, viscous hydrodynamical simulations have suggested that
vortices can be generated at the gap edge only when the kinematic viscosity is small (e.g., $\nu<10^{-5}R^{2}\Omega$, de Val-Borro et al. 2007; Fu \etal 2014).
It is unclear if MHD turbulence has similar effects on vortex formation as the viscous approximation, and 
if AD in a realistic protoplanetary disk can lead to enough suppression of disk turbulence  to facilitate vortex formation at the gap edge.

We will introduce our numerical method
in \S 2. Our results are
presented in \S 3. After a short discussion in \S 4, we will conclude in \S 5.

\section{Method}
\subsection{MHD Simulations}
To study gap opening by planets in disks where turbulence is already fully developed, we restart the global MHD simulations in Zhu \etal (2014b), but with
a 9 $M_{J}$ planet in a circular Keplerian orbit at $r=1$. We then continue the simulations for another 100 orbits. In one
case where a gap edge vortex is present, we continue the simulation until the vortex disappears
around 2000 orbits, as will be
discussed in \S 4.1. 

Since we wish to compare 3-D MHD simulations with  two-dimensional (2-D) viscous hydrodynamical (HD) simulations in 3-D,
we use unstratified disks for 3-D MHD simulations in which the vertical component of gravity is zero. 
We also set the planet potential to be the potential of a line mass (symmetric in the $z$ direction). This
geometry is identical to the symmetry of 2D HD simulations, but allows MHD turbulence to be modeled in full 3D.
To avoid small time steps 
associated with the divergence of a point source potential, 
the planet potential is smoothed with 
a smoothing length
of 0.18, which is close to the Hill radius of the planet.  The detailed form of the smoothed 
planet potential can be found in Zhu \etal (2013).\footnote{We do not include indirect forces
from the planet to star and the disk to star since these second order forces
will not affect the gap opening process. }
Furthermore,
in order to avoid strong disturbance if we suddenly initialize the planet in the disk, we linearly ramp up the 
amplitude of the planet potential in 10 orbits.

The setup of the global MHD simulations is given in detail in Zhu \etal (2014b).
Here we provide a short summary.
The gas dynamics is computed with Athena (Stone \etal 2008), a higher-order
Godunov scheme for hydrodynamics and magnetohydrodynamics using constrained transport
to conserve the divergence-free property for magnetic fields
(Gardiner \& Stone 2005, 2008).  Cylindrical grids (Skinner \& Ostriker 2010) and the particle 
integrator  (Bai \& Stone 2010; Zhu \etal 2014a) are used 
to study the motion of dust particles in global MRI turbulent disks. 
Dust particles are implemented as Lagrangian
particles with the
dust-gas drag term, following
\begin{equation}
\frac{d\mathbf{v}_{i}}{dt}=\mathbf{f}_{i}-\frac{\mathbf{v}_{i}-\mathbf{v}_{g}}{t_{s}}\,,\label{eq:motion}
\end{equation}
where  $\mathbf{v}_{i}$ and $\mathbf{v}_{g}$ are the velocity vectors for particle $i$
and the gas, $\mathbf{f}_{i}$ is the gravitational force experienced by
particle $i$, and $t_{s}$ is the stopping time due to gas drag. 
Different from the setup for gas, we  include the vertical gravitational force for dust particles, in which case dust is
allowed to settle toward the disk midplane. 
The unstratified disk setup without including the vertical gravity is a good approximation
for simulating 
 the gas dynamics at the disk midplane (e.g. within half the gas scale height in the disk, $h=c_{s}/\Omega $),
  since the vertical gravity is 
much smaller than the gas pressure there. However, even close to the disk midplane, dust can 
 have a significant stratification since it can have a much smaller scale height than the gas due to vertical settling.

The initial radial profile of the gas disk in Zhu \etal (2014b) is
\begin{eqnarray}
\rho_{g}(r,\phi,z)=\rho_{g,0}\left(\frac{r}{r_{0}}\right)^{-1} \label{eq:eqinid}\\
T(r,\phi,z)=T_{0}\left(\frac{r}{r_{0}}\right)^{-1/2}\,.\label{eq:eqinit}
\end{eqnarray}
We choose $\rho_{g, 0}=1$, $r_{0}=1$, and $(h/r)_{r=1}=0.1$. The
disk is locally isothermal at each $r$, which is a good approximation in the outer disk (beyond 10s of {\rm AU})
 where the study focused. When we restart the simulation at 100 orbits (where the
orbit is defined as the orbital time at $r=1$, or $2\pi\Omega^{-1}(r=1)$), 
 the surface density of the gas disk is not much different from this 
initial condition since 100 orbits
 is much shorter than the viscous timescale. 

Turbulence is driven by the MRI in both ideal and non-ideal
MHD with ambipolar diffusion (AD). Following Bai \& Stone (2011), we include 
the effect of AD by modifying the induction equation as
\begin{equation}
\frac{\partial {\bmath B}}{\partial t} = \del \times \left({\bmath v} \times {\bmath B}- \frac{4\pi}{c} \eta_{A}{\bmath J}_{\perp} \right),
\end{equation}
where ${\bmath B}$ is the magnetic field, 
 ${\bmath J} _{\perp}= (\del \times {\bmath B})_{\perp}$ is the component of the current density that is
  perpendicular to the direction of the magnetic field, and $\eta_{A}$ is
the ambipolar diffusivity which depends on the disk ionization ratio. 
We choose $\eta_{A}$ to be equal to $v_{A}^{2}/\Omega$, 
where $v_{A}$ is the Alfven speed. This $\eta_{A}$ is  typical for the protoplanetary disk at 10s of {\rm AU} (Bai 2011a). 

For ideal MHD runs, the disk is threaded by
either net vertical  or net toroidal magnetic fields. For AD runs, we only study disks threaded by net vertical fields since
net toroidal fields generate little turbulence in disks dominated by AD.
The initial magnetic fields have a constant plasma $\beta=8\pi \rho c_{s}^{2}/B^{2}$ everywhere in the whole disk.
The initial configuration of magnetic fields is shown in Table 1.  Ideal MHD runs with net vertical and toroidal 
fields have names 
starting with ``V'' and ``T'' respectively, while AD runs start with ``AD''.
For each magnetic field geometry, we have varied the initial plasma $\beta$ by at least one order
of magnitude trying to extend our parameter space of the turbulent stresses.  
Table 1 shows the averaged
stresses  over the region of  $r\times\phi\times z$=[0.95,1.05]$\times$[0,2$\pi$]$\times$[-0.1,0.1].

\begin{table}[ht]
\begin{center}
\caption{Models \label{tab1}}
\begin{tabular}{cccccccc}
\tableline
\tableline
MHD &&&&&& \\
\tableline
\tableline
Run  \tablenotemark{a}   & B  &  $\langle \rho v_{r} \delta v_{\phi} \rangle$/$\langle \rho \rangle$ & $\langle -B_{r}B_{\phi}\rangle$/$\langle \rho \rangle$ & $\alpha$\tablenotemark{b} \\
 & $\beta_{0}$   & ($c_{s}^{2}$) & ($c_{s}^{2}$)    &  \\
\tableline
Ideal MHD &\\
V1e4  & Vert. 1e4 &  7.5e-3 & 0.028  & 0.035  \\
V1e5  & Vert. 1e5 & 5.3e-3 & 0.017  & 0.022  \\
T1e2  & Tor. 100 & 7.5e-3 & 0.024  & 0.032  \\
T1e3   & Tor. 1e3  & 6.2e-3 & 0.018  & 0.025  \\
T1e2H   & Tor. 100& 0.010 & 0.023  & 0.033  \\
\tableline
\multicolumn{2}{l}{Ambipolar diffusion} \\
AD1e3  & Vert. 1e3   & 8.8e-4 & 1.5e-3  & 2.4e-3  \\
AD1e3L \tablenotemark{c}   & Vert. 1e3   & - & 3.5e-4  & -  \\
AD2.5e4   & Vert. 2.5e4 & 5.0e-4 & 1.8e-4  & 6.4e-4  \\
\tableline
\tableline
\tableline
\tableline
Viscous & &&&&& \\
\tableline
\tableline
Run  & $\alpha_{ss}$ & &  &  & &  \\
\tableline
HD023 & 0.023 &&&&& \\
HD016 & 0.016 &&&&& \\
HD0016 & 0.0016 &&&&& \\
HD0002 & 0.0002 &&&&& \\
HD001L & 0.001 &&&&& \\
\tableline
\end{tabular}
\tablenotetext{1}{ All quantities are measured at $r=1$ before inserting the planet.}
\tablenotetext{2}{ $\alpha\equiv\frac{\langle -B_{r}B_{\phi}\rangle}{\langle\rho\rangle c_{s}^2}+\frac{\langle\rho v_{r} \delta v_{\phi}\rangle}{\langle\rho\rangle c_{s}^2}$.}
\tablenotetext{2}{For this case, we measure $\langle -B_{r}B_{\phi}\rangle$ 
at $r=2$ (the gap edge) at 900 planetary orbits after inserting the planet, in order to
represent the disk turbulent strength at the time when the vortex disappears.
The planetary torque affects $\langle \rho v_{r} \delta v_{\phi} \rangle$ significantly. }
\end{center}
\end{table}

Our cylindrical grids span from 0.5 to 4 in the $r$ direction, $0$ to $2\pi$ in the $\phi$ direction, and 
-0.1 to 0.1 in the $z$ direction. 
The grid is uniformly spaced in all $r$, $\phi$ and $z$ directions. Our standard simulations
have the resolution of $576\times1024\times32$ in the $r$, $\phi$, and $z$ directions, which is 16 grids per $h$ at $r=1$
 in all three directions. 
As shown in Zhu \etal (2014b), this standard resolution is fully converged for MRI turbulence in all our cases. On the other hand,
we run one high resolution simulation (T1e2H with $1152\times2048\times64$) to 70 orbits after inserting the planet to make sure
 resolution does not affect vortex generation at the gap edge.

At the outer radial boundary, the physical quantities in ghost zones 
are set to be fixed at the initial values (as in Zhu \etal 2014a).
At the inner boundary, the open boundary condition used in Sorathia \etal (2012) has been applied to allow
mass accretion.
Periodic boundary conditions have been applied in both $\phi$ and $z$ directions.

\begin{figure}[ht!]
\centering
\includegraphics[width=0.5\textwidth]{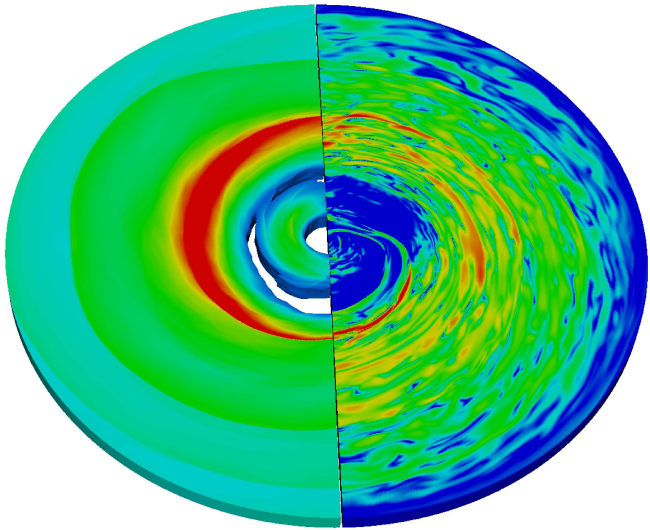} 
\vspace{-0.3 cm}
\caption{Contours for the gas density (the left half disk) and $B^{2}$(the right half disk) in the AD1e3 run.
The movie can be downloaded from the online material of this paper. 
} \label{fig:dB2}
\end{figure}

Volume rendering of the gas properties in simulation AD1e3 is shown in Figures \ref{fig:dB2}.

We evolve seven types/species of particles simultaneously with the gas. 
For each particle type,  there are
$10^{6}$ particles.
We assume that
all these particles are in the Epstein regime, so that
 the dust stopping time (Whipple 1972; Weidenschilling 1977, we use the notation from Takeuchi 
\& Lin 2002) is
\begin{equation}
t_{s}=\frac{ s \rho_{p} }{\rho_{g}v_{T}}\,,\label{eq:ts}
\end{equation}
where $\rho_{g}$ is the gas density, s is the dust particle radius, $\rho_{p}$ is the dust particle density (we 
choose $\rho_{p}$=1 g cm$^{-3}$), $v_{T}$=$\sqrt{8/\pi}c_{s}$, and $c_{s}$ is the gas sound speed. The 
dimensionless form of $t_{s}$ is $T_{s}\equiv t_{s}\Omega$.

Each particle type in our simulations has one certain size ($s$). 
The dust stopping time for each particle type at $r=1$ 
in the initial condition  is given in Table 2.
Given a realistic protoplanetary disk structure, particles in our simulations
can be translated to real particles having some physical sizes  in protoplanetary disks..
Using Equation (\ref{eq:ts}) and $c_{s}=h\Omega$, 
a dust particle with size $s$ at the midplane of a realistic disk ($\rho_{mid}=\Sigma/(\sqrt{2\pi}h)$) has
\begin{equation}
T_{s}=\frac{\pi s \rho_{p}}{2\Sigma_{g}}=1.57\times10^{-3}\frac{\rho_{p}}{1 {\rm g \, cm^{-3}}}\frac{s}{1\, {\rm mm}}\frac{100\, {\rm g \,cm^{-2}}}{\Sigma_{g}}\,.\label{eq:ts3}
\end{equation}
Thus, for different realistic disk structures, Table 2 also gives the real particle sizes that our simulated particle types correspond to.
To relate our results  with the latest ALMA observation on Oph IRS 48 (van der Marel et al. 2013, Bruderer \etal 2014) and HD 142527 (Casassus \etal 2013), 
we give the corresponding size of particles in
Oph IRS 48 and HD 142527 in last two columns.  
However, with a realistic surface density,
some particle types are in the Stokes rather than the Epstein regime. Those particle types can
only be considered as a numerical experiment to explore the effect of a large stopping time on the particle distribution,
rather than a realistic model of such particles.

\begin{table}[ht]
\begin{center}
\caption{Corresponding particle sizes in different disks \label{tab1}}
\begin{tabular}{lccccc}
 
\tableline\tableline
Par  & $t_{s}\Omega$    &   Planet at  &   Planet at       &  Planet at         & Planet at     \\
                        &  At $r$=1     &    5 AU in &   20 AU in     &  30 AU  in        & 70 AU in    \\
                        & Initially  &  $\alpha$ Disk \tablenotemark{a} & $\alpha$ Disk & Oph IRS 48 \tablenotemark{b} & HD 142527  \tablenotemark{c}\\
\tableline
a &   0.007041   & 1.5 mm & 0.4 mm & 3 $\mu$m & 14 $\mu$m \\
b &   0.07041   & 1.5 cm &  4 mm  &  30 $\mu$m  & 140 $\mu$m\\
c &   0.7041 &  15 cm & 4 cm  & 0.3 mm & 1.4 mm\\
d &   7.041   & Stokes\tablenotemark{d} & 40 cm   & 3 mm & 1.4 cm\\
e  &   70.41  &  Stokes  & 4 m & 3 cm  & 14 cm\\
f  &   704.1  &  Stokes  & Stokes & 30 cm & 1.4 m\\
g &   7041  &  Stokes  & Stokes & 3 m  &  14 m\\
\tableline
\end{tabular}
\tablenotetext{1}{ The $\alpha$ disk has $\Sigma_{g}=178 (r/{\rm AU})^{-1}$ g cm$^{-2}$, which is the surface density of a constant $\alpha=0.01$ accretion disk with
$\dot{M}=10^{-8}\msunyr$,
and $ T = 221 (r/{\rm AU})^{-1/2}$~K }
\tablenotetext{2}{ Oph IRS 48 disk with a 2 M$_{\odot}$ central star,  
$\Sigma_{g}=1.92 (r/{\rm AU})^{-1}$ g cm$^{-2}$ and 
$ T = 542 (r/{\rm AU})^{-1/2}$~K (Bruderer \etal 2014).}
\tablenotetext{3}{ HD 142527 disk, $\Sigma_{g}=22 (r/{\rm AU})^{-1}$ g cm$^{-2}$ and 
$ T = 354 (r/{\rm AU})^{-1/2}$~K (Casassus \etal 2013). }
\tablenotetext{4}{ Particles in the Stokes regime, see text.} 
\end{center}
\end{table}

Initially, we set the dust having the same surface density as the gas, and all the particles are
placed at the disk midplane. When we restart our simulations at 
100 orbits, most of our particle types have already reached
to vertical equilibrium states since the settling timescale of dust particles ($\Omega^{-1}(T_{s}+T_{s}^{-1})$) is only
22 orbits at $r$=1 for the smallest particles in our simulations (Par. a). 
However, during the time that small particles with $T_{s}\ll1$ (e.g., Par. a) reach a vertical equilibrium, 
big particles with $T_{s}\sim 1$ (e.g., Par. c and Par. d) have significant radial drift. 
The dust surface density at 100 orbits when we restart the simulations before inserting the 
planet can be found in Figure 3 in Zhu \etal (2014b).

\subsection{Viscous Hydrodynamical Simulations}
We have also carried
out 2-D hydrodynamical(HD) simulations using viscosity to approximate turbulence
to study particle trapping at the gap edge and vortices. 
The kinematic viscosity in these simulations have the same
$r-\phi$ stresses as those in the MHD simulations.
Note that in viscous HD disks the $r-\phi$ component of the viscous 
stress tensor
 $T_{v}\sim\nu\rho (r \partial (v_{\phi}/r )/\partial r + 1/r \partial v_{r}/\partial \phi)\sim \nu\rho 1.5 \Omega$. 
Equating $T_{v}$ with the MRI stress $T_{MRI}\sim\alpha\rho c_{s}^{2}$ leads 
to $\nu=\alpha c_{s}^2/1.5 \Omega=\alpha_{ss} c_{s}^2/ \Omega$, where $\alpha_{ss}=\alpha/1.5$ as
 in Shakura \& Sunyaev (1973). Since $\alpha \propto r^{-5/4}$ and $c_{s}^{2}\propto r^{-1/2}$ in our MHD simulations
(Zhu \etal 2014b), we have $\nu=(\alpha_{0}/1.5)c_{s,0}^{2}\Omega_{0}^{-1}  (r/r_{0})^{-1/4}$,
where the subscript $0$ denotes the quantities at $r=r_{0}$.
For example, in order to compare with  
the MHD case V1e4 which has $\alpha_{0}$=0.035, the corresponding viscous run
 has $\nu=0.023 c_{s,0}^{2} \Omega_{0}^{-1}(r/r_{0})^{-1/4}$, which is denoted as
 run HD023.

All HD simulations with their $\alpha_{ss}$ at $r=1$ are shown in Table 1.
Overall, run HD023 corresponds to both V1e4 and T1e2, since these two MHD cases have
similar $\alpha$. HD016 corresponds to both V1e5 and T1e3. HD0016
corresponds to AD1e3, and HD0002
corresponds to AD2.5e4. \footnote{In AD2.5e4, the Reynolds stress fluctuates significantly along the $r$ direction. 
We find $\alpha_{ss}=0.0002$ roughly approximates the averaged stress.}

Considering viscous simulations converge more easily than MHD simulations,
our 2D viscous disks have a resolution of $288\times 512$ in the $r$ and $\phi$ directions.
This is equivalent to 8 grid cells per $h$ at $r$=1, which
is adequate to study gap edge vortices, as in our previous inviscid hydrodynamical 
simulations (Zhu \etal 2014a). Other quantities are initialized as in our MHD simulations. There
are seven types of particles in the simulation. For each type, there are $10^{6}$ particles initially.  We run
these HD simulations for 100 orbits without planets, then insert a 9 $M_{J}$ planet at $r=1$, and continue the simulations for another
100 orbits.

\section{Results}
\subsection{Gas Disks}

\begin{figure*}[ht!]
\centering
\includegraphics[trim=0cm 2.8cm 0cm 0cm, width=1.0\textwidth]{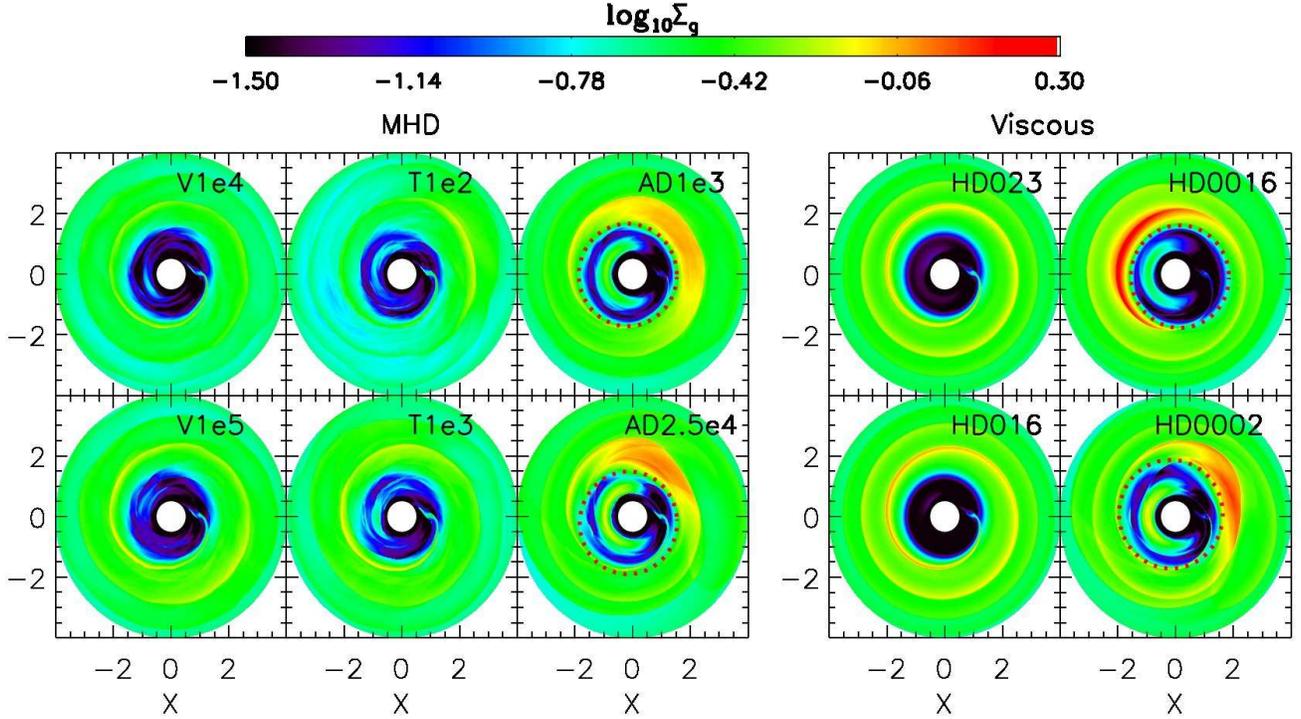} 
\vspace{-0.3 cm}
\caption{Gas disk surface density at 100 planetary orbits after inserting the planet for MHD runs (left panels) 
and corresponding viscous hydrodynamical runs (right panels). The HD023 run corresponds
to both V1e4 and T1e2 runs, and the HD016 run corresponds to both V1e5 and T1e3 runs. 
The HD0016 run
corresponds to AD1e3, while HD0002 corresponds to AD2.5e4. 
In MRI turbulent disks, the
vortex is generated  only when AD dominates in the disk (AD runs). 
For generating the vortex at the gap edge, the equivalent viscosity parameter $\alpha$ needs to be
 $\lesssim10^{-3}$, similarly in both MHD and viscous HD runs. The red dotted curves are ellipses
 to fit the shape of the gap edges.
} \label{fig:imagegas2}
\end{figure*}

\begin{figure*}[ht!]
\centering
\includegraphics[trim=0cm 2.8cm 0cm 0.4cm, width=1.0\textwidth]{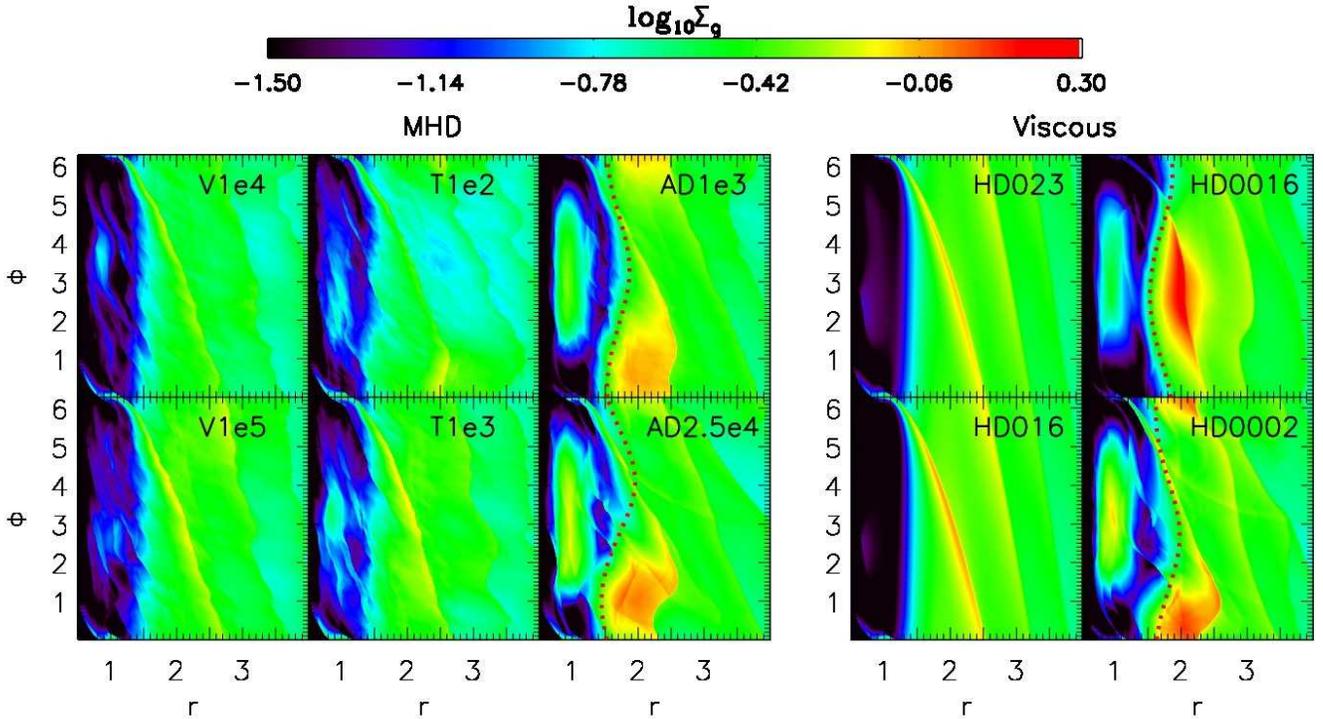} 
\vspace{-0.3 cm}
\caption{Similar to Fig. \ref{fig:imagegas2}, but in the $r-\phi$ plane.
} \label{fig:rphi}
\end{figure*}

{The gas surface density for both MHD and viscous HD runs at the end of the simulations 
is shown in Figures \ref{fig:imagegas2} and \ref{fig:rphi}.
The most prominent asymmetric feature is the gap edge vortex in the MHD simulations 
where AD dominates or in the viscous HD simulations where the viscosity is low.} 
In ideal MHD runs, turbulence is so strong (the equivalent $\alpha\sim 10^{-2}$) that the vortex cannot form at the gap edge. 
With AD, turbulence is significantly suppressed ($\alpha\lesssim 10^{-3}$), and the vortex forms at the gap edge. 
Similarly in corresponding HD runs, 
the vortex only forms when the viscosity is small  (HD0016 and HD0002).
For runs with a vortex at the gap edge,  the vortex is more elongated in MHD runs that have 
stronger turbulence (AD1e3) or in HD runs with a larger viscosity (HD0016). Overall, in MRI turbulent disks, 
the vortex is only generated when turbulence is weak, and the similarity
between MHD runs and viscous HD runs implies that, in future, we may use HD simulations with
viscous approximation to study the production of gap edge vortices in turbulent protoplanetary disks.

{ Another asymmetric gas feature is the ``C'' shaped horseshoe region within the gap.}
This region is more prominent in MHD runs with less turbulence or in HD runs with lower viscosities. 
This trend can be explained by the gap opening theory including non-linear wave
 propagation (Goodman 
\& Rafikov 2001), which states that 
 density wakes excited by the planet propagate and shock in disks, and two gaps are 
opened at the shocking distance on both sides of the planet. Stronger
turbulence or a higher viscosity depletes the horseshoe region into these two gaps by viscous diffusion, leaving
a less prominent  ``C'' shaped horseshoe region.
One noticeable difference at the horseshoe region between MHD and HD runs is that the horseshoe
region still exists even in highly turbulent ideal MHD runs, while it disappears in the corresponding viscous runs. 
In ideal MHD runs, the horseshoe region is also more prominent in disks threaded by net toroidal fields (T1e2 and T1e3) than in disks threaded by net vertical
fields (V1e4 and V1e5), suggesting that the magnetic field geometry affects the gap opening process. This is consistent
with gap opening studies using MHD shearing box simulations (Zhu \etal 2013), which found that
the corotation region of MHD disks threaded by net vertical fields is more depleted than those
threaded by net toroidal fields.

{ The gap edge also appears eccentric due to the presence of the large vortex there.} This is more clearly shown
in Figure \ref{fig:rphi}. The gap edges can be fit with ellipses having eccentricities of 0.1-0.2 (red dotted curves). More discussion
on the shape of the gap edge is given in \S 4.2.

{ The spiral wakes look different between MHD and HD runs.} The spiral wakes excited by the planet can be disrupted by turbulence in MHD runs, 
especially in runs where the turbulence is strong (e.g., ideal MHD runs) or in regions where the wakes are weak (e.g., at large r), while the spiral wakes are always
smooth and continuous  in viscous HD runs. 

\subsection{Dust Disks}

\begin{figure*}[ht!]
\centering
\includegraphics[trim=0cm 2.8cm 0cm 0.4cm, width=1.0\textwidth]{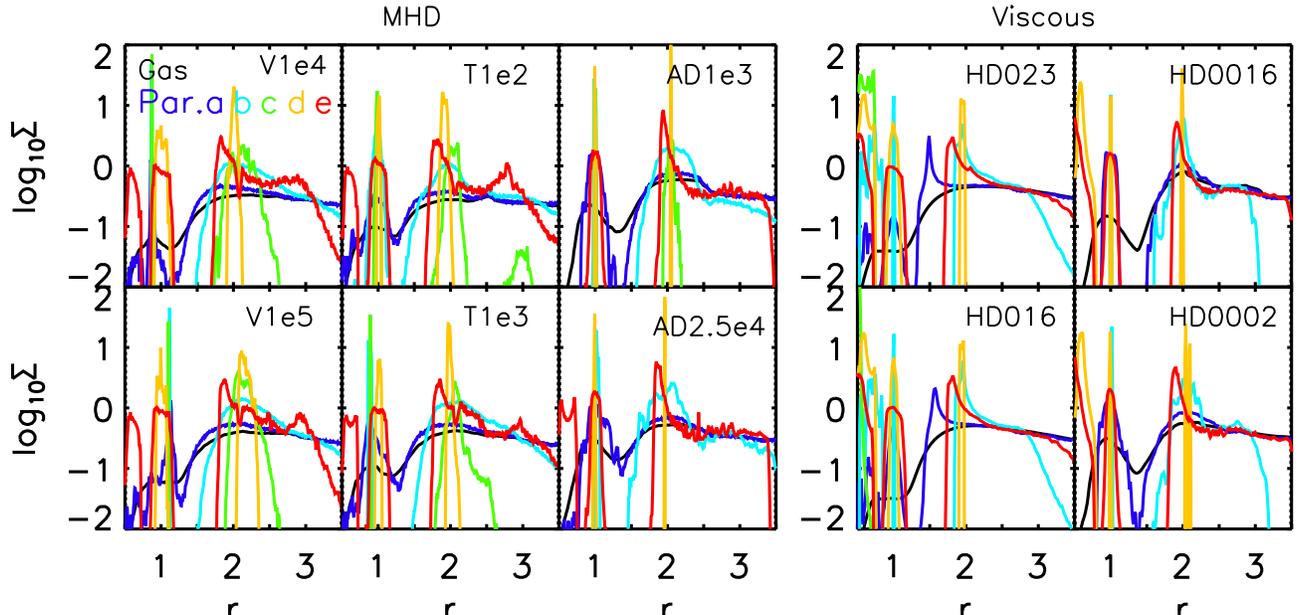} 
\vspace{-0.3 cm}
\caption{Vertically and azimuthally averaged 
disk surface density for both gas and dust in MHD runs (left panels) and viscous runs (right panels) at 100 planetary orbits after inserting the planet.
} \label{fig:oned}
\end{figure*}

\begin{figure*}[ht!]
\centering
\includegraphics[trim=0cm 0.1cm 0cm 0.4cm, width=0.8\textwidth]{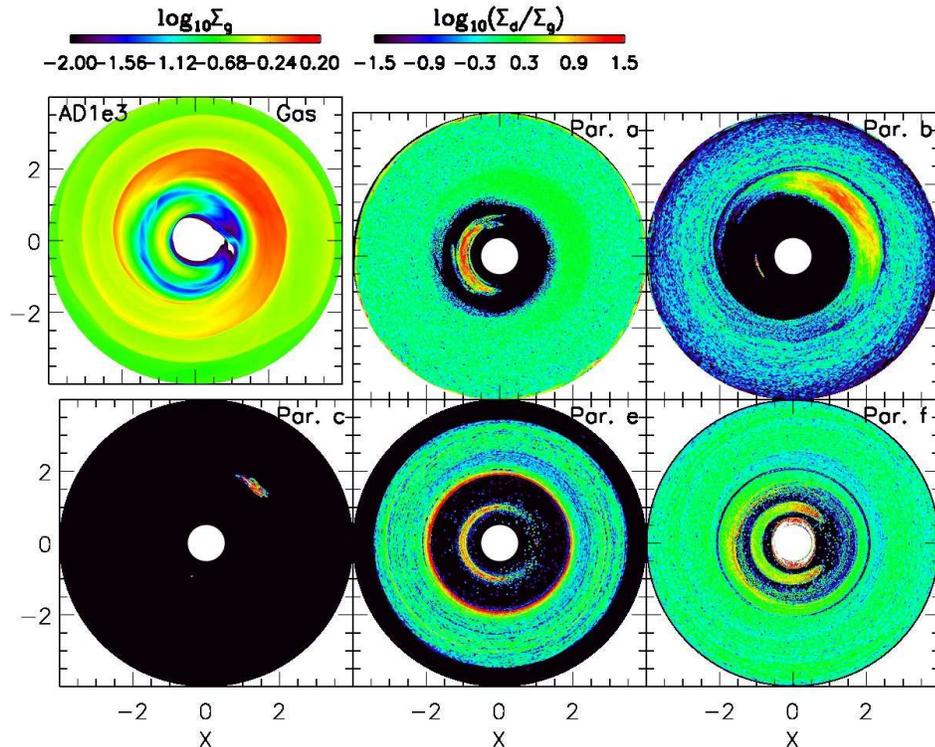} 
\vspace{-0.3 cm}
\caption{Disk surface density at 100 planetary orbits after inserting the planet for AD1e3.
The upper left panel shows the surface density of the gas disk, while
the other five panels show the ratio between the dust surface density and the gas surface density.
With the increasing particle size from Par. a to Par. c (or 3 $\mu$m to 0.3 mm in the disk around Oph IRS 48), 
dust concentrates more in 
the vortex.  Par. e having $T_{s}\gg1$ start to decouple from the gas and the concentration
at the gap edge is in a more axisymmetric fashion. 
Par. f exhibit additional gaps at mean motion resonances of the planet.
} \label{fig:imagegaspar}
\end{figure*}

\begin{figure*}[ht!]
\centering
\includegraphics[trim=0cm 2.8cm 0cm 0.2cm, width=1.0\textwidth]{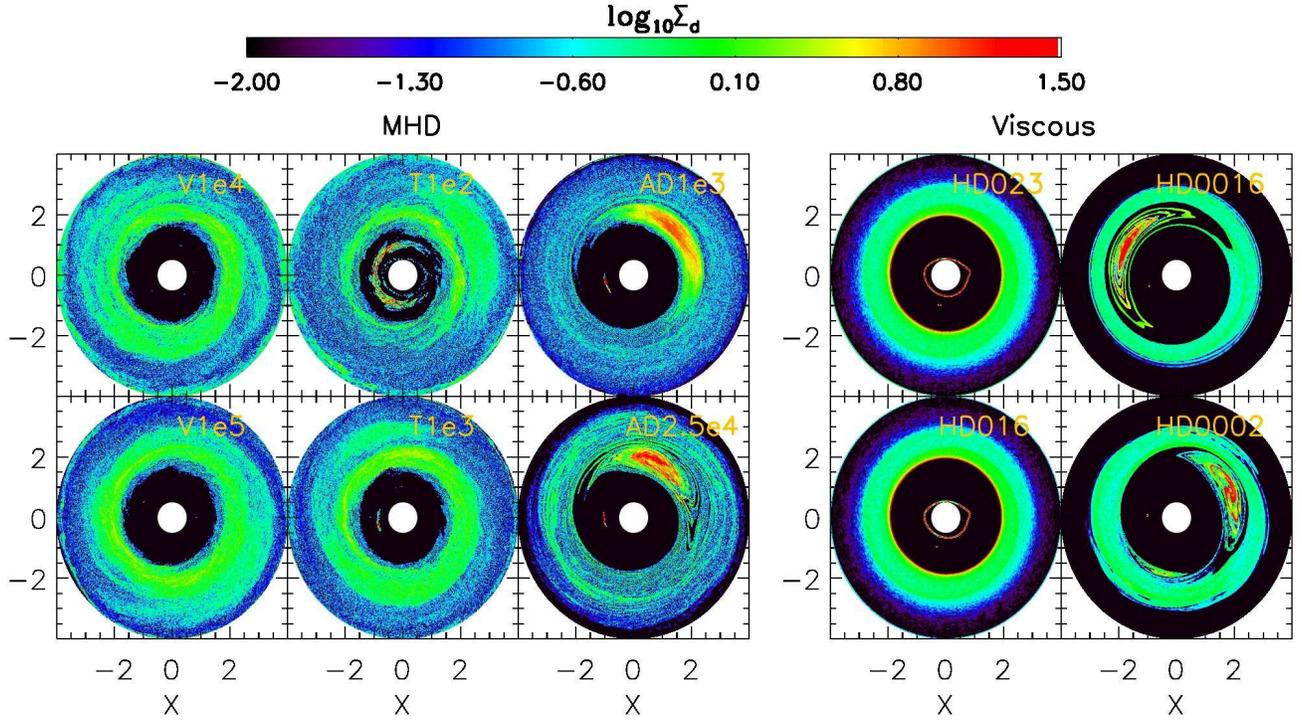} 
\vspace{-0.3 cm}
\caption{Similar to Figure \ref{fig:imagegas2}, but shows the surface density of 
Par. b.} \label{fig:imagepar}
\end{figure*}

\begin{figure*}[ht!]
\centering
\includegraphics[trim=0cm 1.8cm 0cm 1.9cm, width=1.0\textwidth]{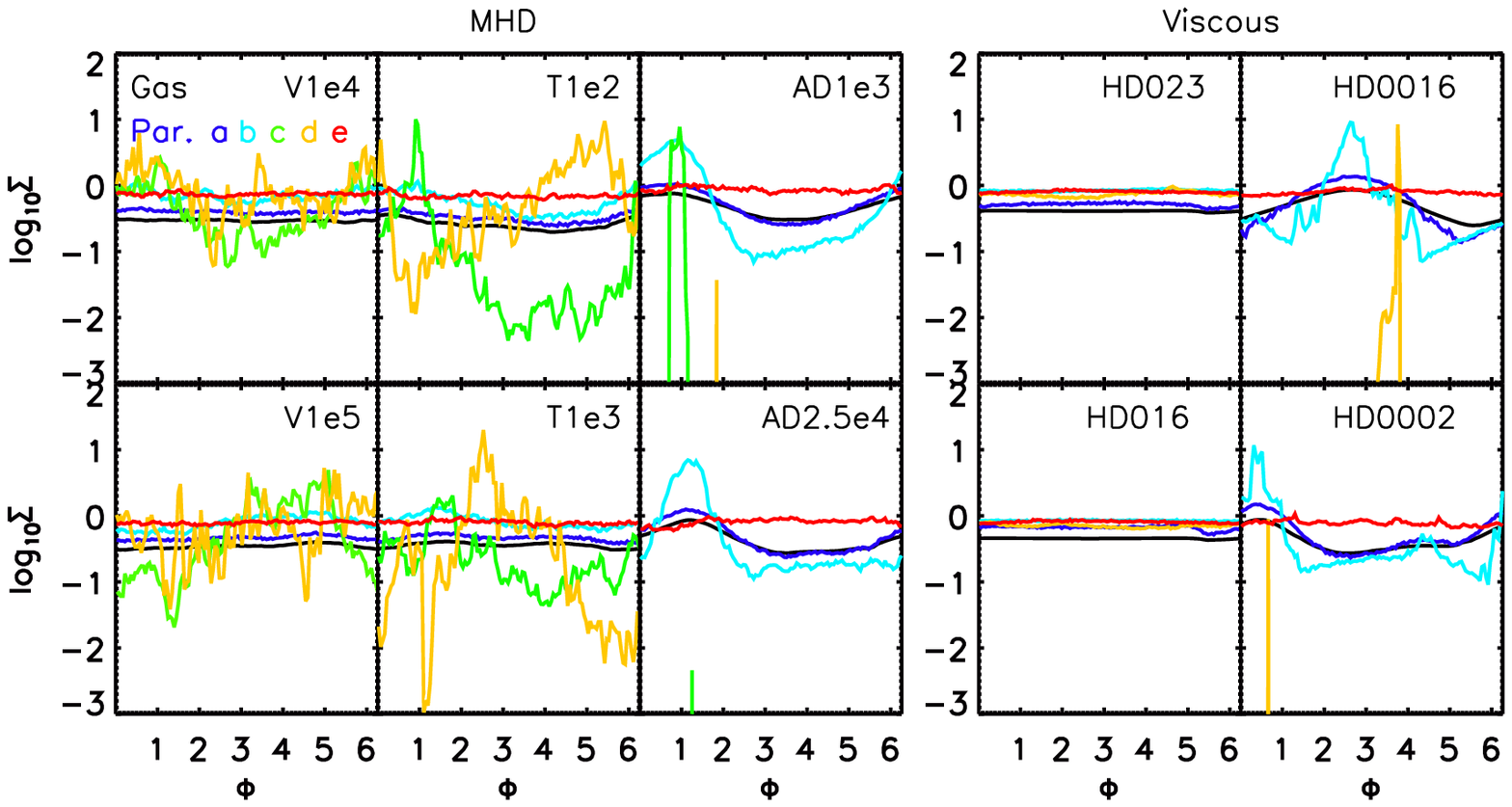} 
\vspace{-0.8 cm}
\caption{Gas (black curves) and particle (colored curves) surface density at the gap edge ($r\sim$2) along the $\phi$ direction for the MHD (left panels) and viscous HD runs (right panels) at 100 planetary orbits after inserting the planet. 
The surface density shown is derived by averaging the disk surface density from $r$=1.5 to $r$=2.5.  Par. b, c, and d have significant increase of the surface 
density within the vortex in AD1e3,
AD2.5e4, HD0016, and HD0002.  Par. c is absent in AD2.5e4 and all viscous HD runs since
these particles have all drifted within $r=1$ before the gap starts to develop.  In ideal MHD runs (V1e4, V1e5, T1e2, T1e3), Par. c, d with $T_{s}\sim 1$ show
 variations even when there is no vortex at the gap edge. } \label{fig:cut2D}
\end{figure*}

{ Figure \ref{fig:oned} shows the vertically and azimuthally averaged 
disk surface density for both gas and dust at the end of all our runs.} Particles are trapped
at the gap edge ($r\sim 2$) and the horseshoe region ($r\sim1$). The smallest particles (Par. a) 
couple with gas very well and their density only increases slightly at
the gap edge. With increasing particle size, trapping at the gap edge is more apparent. The maximum dust 
concentration at the gap edge
occurs for particles with $T_{s}\sim 1$ (e.g., Par. c and d).\footnote{Par. c are not present in AD2.5e4 and viscous runs. 
This is because when we restart the simulations at 100 orbits, Par. c have drifted to $r<1$ in these runs.} It is also noticeable that, for particles that have
$T_{s}$ closer to 1, the gap edge
moves outward and the gap becomes bigger. For example, the gap edge of Par. c is at $r\sim 2$, compared to the gap edge
of gas and Par. a at $r\sim 1.5$. 

{ Since vortices at the gap edge can trap dust particles, 
dust disks have stronger asymmetric features at the gap edge than the gas disk has, as shown in Figure \ref{fig:imagegaspar}.}
Take Par. b as an example, its surface density within the vortex can be $\sim$10 times higher than outside. The crescent-shaped vortex region
with a significant Par. b concentration also has a large radial width: $\Delta r\sim 1$ at $r=2$ compared with $h\sim 0.24$ at $r=2$.
When particles have $T_{s}\sim$1 (e.g., Par. c), almost all the particles have been trapped at the vortex center.  Particles having 
$T_{s}> 1$ (e.g., Par. d and Par. e) start to decouple from the gas and they take longer to drift to the
vortex center. For even larger particles (e.g., Par. f), gaps are opened at mean motion resonances of the planet,
similar to Zhu \etal (2014a).  

{ To compare dust distribution between MHD runs and HD runs,
Figure \ref{fig:imagepar} shows
 the surface density contours of Par. b in all the runs. 
Clearly, the dust distribution is quite similar between MHD runs
and the corresponding HD runs.} 
Par. b are distributed almost axisymmetrically in ideal MHD runs which have high turbulent levels,
 or in corresponding HD runs with high viscosities. The dust is trapped in the vortex in less turbulent (AD runs) or corresponding less viscous runs.
 The particle distribution in viscous HD runs has sharper edges, since there is no turbulent diffusion for particles in these cases. 

{  To quantify the asymmetry of dust distribution at the gap edge, we average
the disk surface density from 
$r=1.5$ to $2.5$ and plot the result along the $\phi$ direction in Figure \ref{fig:cut2D}.}
In ideal MHD runs which have turbulent levels of $\alpha\gtrsim0.01$, 
Par. c and d (with
$T_{s}\sim 1$) are distributed quite asymmetrically at the gap edge, even though the gas distribution is relatively uniform
there. The distributions of these particles look like sinusoidal functions in the $\phi$ direction over $2\pi$,
and the amplitude of the variation is slightly larger in disks threaded by net toroidal fields (T1e2 and T1e3) than in disks threaded by net 
vertical fields (V1e4 and V1e5). On the other hand,
Par. c and Par. d peak at different azimuthal positions, suggesting that they are affected by transient
perturbations probably due to MRI turbulence
instead of persistent large scale gas features which concentrate particles at the same position. 
Other particle types having  $T_{s}\not\approx 1$ (Par. a, b, and e)
are distributed quite axisymmetrically at the gap edge in the ideal MHD runs.
For these particles,
the surface density variation in the $\phi$ direction is less than a factor of 2.

{ In AD runs, the gap edge vortex leads to a moderate surface
density variation for the gas, but a dramatic increase of the dust surface density within the vortex.}
Figure \ref{fig:cut2D} shows that the gas surface density within the vortex can be $\sim$ 4 times of the value outside. 
However, for Par. b having $T_{s}\sim 0.07$, the surface density within the vortex can be 100 times larger
than outside. The concentration of Par. a within the vortex is also noticeable. 
The vortex can efficiently trap dust particles that span 3 orders of magnitude in size (Par. b, c, and d)
within 100 planetary orbits, producing more than a factor of 10
enhancement in the dust surface density.

{  Figure \ref{fig:cut2D} also shows that, in viscous HD runs the particle distribution at the gap edge is similar to that in MRI turbulent disks, except
for particles with $T_{s}\sim 1$.} Par. d in HD023 and HD016 have a relatively flat distribution
at the gap edge, while they have large variations in the corresponding ideal MHD runs. This again suggests
that the distribution of particles with $T_{s}\sim 1$ in MRI turbulent disks
is affected by transient turbulent fluctuations. 

\section{Discussion}
\subsection{Lifetime of the Vortex}
\begin{figure*}[ht!]
\centering
\includegraphics[width=0.95\textwidth]{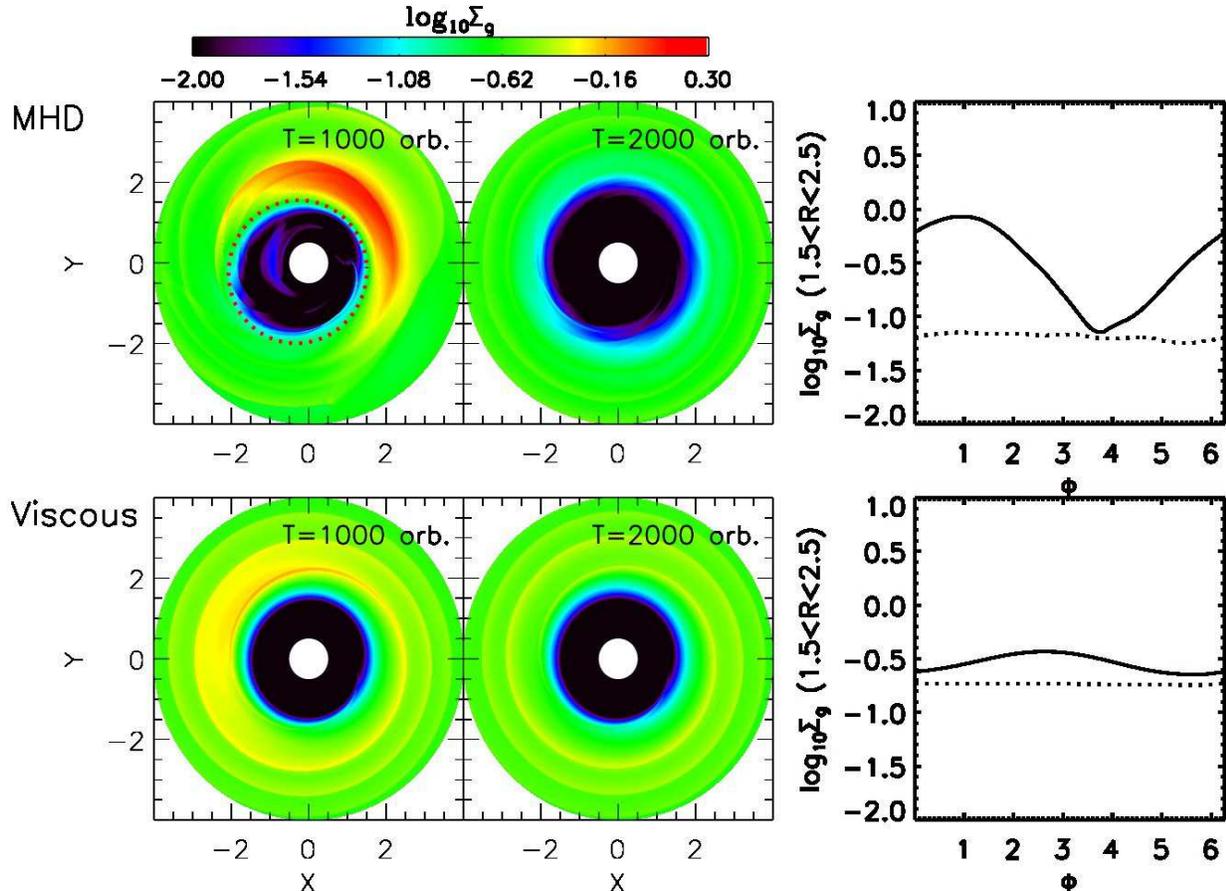} 
\vspace{-0.3 cm}
\caption{Left and middle panels: the gas disk surface density at 1000 (left panels) and 2000 (middle panels) planetary orbits
for the MHD (AD1e3L: upper panels) and the viscous HD runs (HD001L: lower panels). Right panels:
the gas surface density at the gap edge (averaged from $r$=1.5 to $r$=2.5) along the $\phi$ direction for the MHD 
(the upper panel) and the viscous runs (the lower panel) at 1000 (solid curves) and 
2000 planetary orbits (dotted curves). The vortex suffers significant dissipation in the viscous HD run at 1000 orbits.
At 2000 orbits, the vortex disappears in both the MHD and viscous HD runs. 
} \label{fig:imagegaslong}
\end{figure*}

{  The longer the vortex persists, the higher chance it has of being observed.}
To see if the gap edge vortex persists for much longer than 100 orbits, 
we repeat our MHD simulation AD1e3 but run it
up to 2000 orbits (denoted as AD1e3L). To expedite the calculation, we do not include
dust particles. For comparison,
we  have also carried out the corresponding HD simulation for 2000 orbits that uses 
viscosity to approximate turbulence (HD001L).

{ The resulting disk surface density at 1000 and 2000 orbits is shown in Figure \ref{fig:imagegaslong},
which demonstrates that the gap edge vortex can persist for 1000 orbits (left panels) in both MHD simulations and viscous HD simulations.} 
There are some subtle
differences between MHD and HD runs, that is the vortex in the MHD 
simulation is as strong as it was at 100 orbits but the vortex  in the viscous run becomes significantly weaker with time.
 However, at 2000 orbits (middle panels),  the gap edge vortex disappears in both cases. 
To make sure that there are no weak vortices at 2000 orbits that can concentrate dust particles, 
we inject dust particles in the turbulent disk at 2000 orbits,  continue
running the simulation for 20 orbits, and confirm that dust distribution is quite axisymmetric at the end of the simulation. 

\subsection{Eccentric Gap Edge}
{ Due to the existence of the large vortex at the gap edge (Figures \ref{fig:imagegas2} and \ref{fig:rphi}),
 the gap edge appears eccentric.} We can roughly fit
the gap edge with an ellipse. One focus of the ellipse is at the central star,
 and the periapsis of the ellipse is at the vortex. 
The eccentricities ($e$) of our fitted ellipses in Figures \ref{fig:imagegas2} and \ref{fig:rphi}
 is 0.1 in AD1e3, 0.15 in AD2.5e4, and 0.1 in both HD0016, and
HD0002. At 1000 orbits, the fitted ellipse for AD1e3L (Figure \ref{fig:imagegaslong})
has an eccentricity of 0.2. 

{ We want to emphasize that the  gap edge appears eccentric only  due to
 the presence of the vortex.} 
Fluid elements at the gap edge do not necessarily move along eccentric streamlines, in contrast to motion in eccentric disks that are dynamically excited under certain disk conditions by 
eccentric Lindblad resonances  (Lubow 1991, Kley \& Dirksen 2006).
The eccentric gap edge precesses at very different rates in these two cases.
 The vortex orbits around the central star at the local
Keplerian speed (Zhu \etal 2014a), so that the resulting elliptical gap edge is precessing 
at the Keplerian speed in the inertial frame. In an eccentric disk due to eccentric Lindblad resonances,  
the fluid elements at the gap edge
follow elliptical orbits and the eccentric gap edge is almost stationary in the inertial frame (or
 has a very small precession rate), which has inefficient dust trapping at the gap edge (Hsieh 
\& Gu 2012, Ataiee \etal 2013).

\subsection{Synthesizing ALMA observations for Oph IRS 48 and HD  142527}

\begin{figure*}[ht!]
\centering
\includegraphics[width=1.\textwidth]{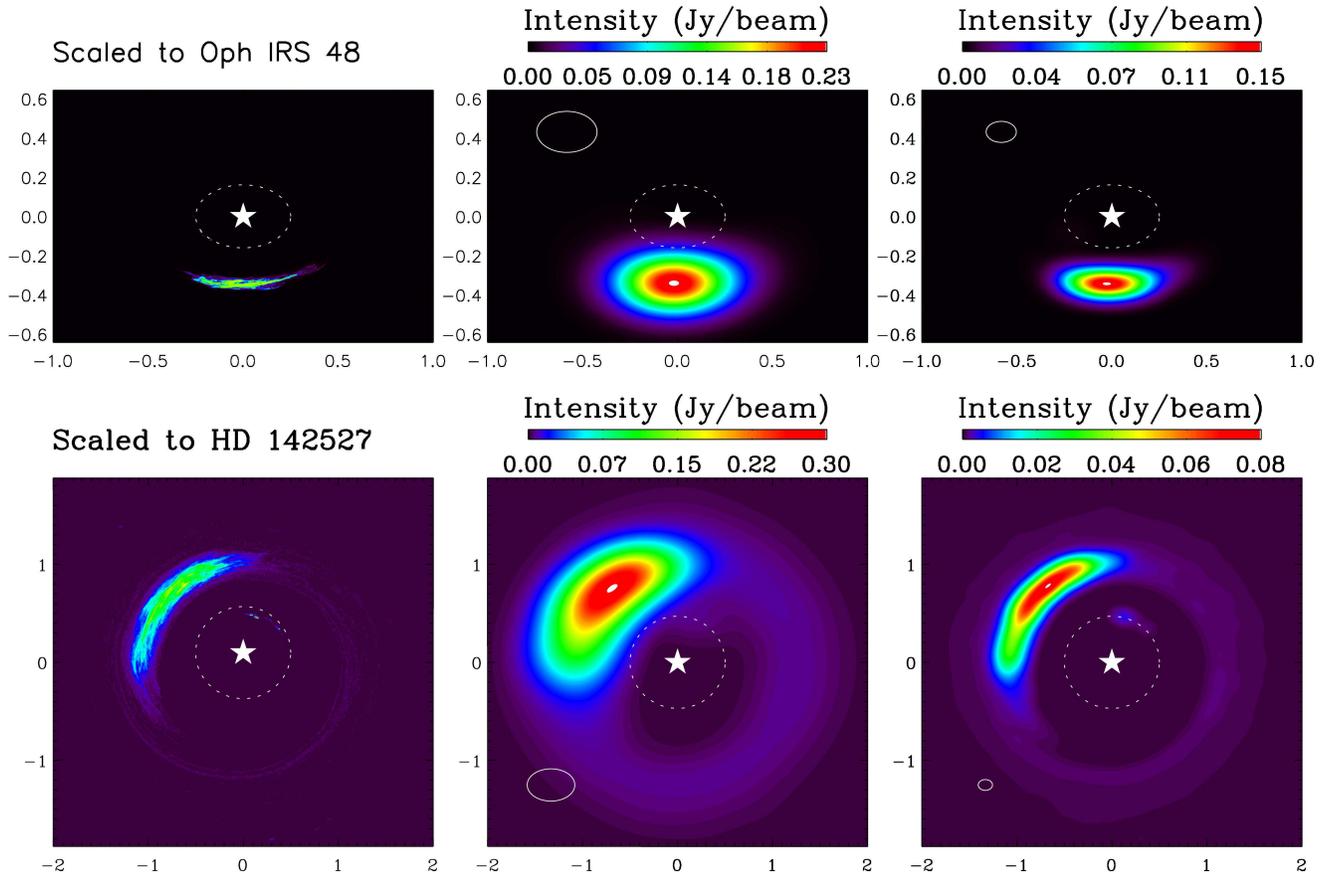} 
\vspace{-0.3 cm}
\caption{Synthetic ALMA observations for Oph IRS 48 (upper panels) and HD 142527 (lower panels) based on our MHD simulation with ambipolar diffusion.
The leftmost panels show the intensity while the middle panel convolves the intensity with the radio beam size given by van der Marel \etal (2013) and Casassus \etal (2013).
By comparing with Figure 1 in van der Marel \etal (2013) and Casassus \etal (2013), good agreements have been achieved. 
The rightmost panels show the prediction for ALMA cycle 2 observations with the radio beam size of 0.16"$\times$0.11".  Thinner lopsided pattern
and horseshoe region may be revealed by ALMA cycle 2 observations.
} \label{fig:ALMA}
\end{figure*}

The gap edge vortex  in AD dominated simulations is promising
to explain disk asymmetry revealed by ALMA observations. 
In this section, we use dust distribution in our simulations to produce synthetic
ALMA observations for Oph IRS 48 and HD 142527.

To better sample the dust size distribution, 
we  run another MHD simulation which has the same parameters as MHD case AD1e3
except that Par. a to g at $r$=1 have stopping times of [0.007, 0.022, 0.07, 0.22, 0.7, 2.2, 7] instead of [0.007, 0.07, 0.7, 7, 70, 704, 7041]. 
Thus, for Oph IRS 48, Par. a to g correspond to grains with radii of 3, 10, 30, 100, 300, 1000, 3000 $\mu$m.
In this case, the ratio between sizes of two adjacent particle types is only $\sim$3 instead of 10. 
After running the simulation without planets for 100 orbits, we inject the planet and run the simulation
for another 200 orbits.

To scale the dust surface density from simulations to the dust surface density in Oph IR 48, we
assume that the dust in Oph IRS 48 has a power law distribution n(s)$\sim s^{-3.5}$ from 0.005 $\mu$m
to 550 $\mu$m. We assume that
dust smaller than 1.7$\mu$m follows the gas distribution (which is justified by good coupling between small dust and gas).
Particles in size bins of [1.7$\mu$m, 5.5$\mu$m], [5.5$\mu$m, 17$\mu$m], [17$\mu$m, 55$\mu$m],  
[55$\mu$m, 170$\mu$m], and [170$\mu$m, 550$\mu$m]  are represented by the surface density distributions of 
Par. a, Par. b, Par. c, Par. d and Par. e in our simulations. Then  by assuming
 the total dust to gas mass ratio of 1:30 \footnote{Bruderer \etal 2014 suggests that the dust-to-gas mass ratio is 1:10 at 60 AU. However, we find
 that the center of the vortex is optically thick already with dust-to-gas mass ratio of 1:30. Higher dust-to-gas mass ratio
won't affect the results much.} and the gas surface density of 
$\Sigma_{g}=1.92 (r/{\rm AU})^{-1}$ g cm$^{-2}$, we can scale our simulation to derive the surface density of dust in each size bin
for Oph IRS 48.
 
To translate the surface density to intensity, we need to know the opacity for dust in each size bin.
We use Mie theory to calculate the dust opacity at 750 GHz (400 $mu$m) for grains with the radii of
 3, 10, 30, 100, 300 $\mu$m. We assume that the grain opacity is dominated by amorphous (olivine) silicates (Dorschner \etal 1995\footnote{The optical constants are given in
 http://www.mpia-hd.mpg.de/HJPDOC/PAPERS/DORSCHN.95/olmg50.ric.})  and the abundance of amorphous silicate in grains is 34$\%$. The dust
 opacities (per unit dust mass) at 750 GHz for grains with radii of  3, 10, 30, 100, 300 $\mu$m are 0.49, 0.58, 4.1, 33, 6.7 cm$^{2}$ g$^{-1}$. 
 
Finally, we multiply the surface density of dust in each size bin with the corresponding opacity and add
these products together to derive the total optical depth of the disk. With the total optical depth, the temperature
distribution of $ T = 542 (r/{\rm AU})^{-1/2}$~K, the distance of 180 pc, 50$^{o}$ inclination, and the ALMA beam size of 0.32"$\times$0.21",
we can synthesize ALMA observations for Oph IRS 48 based on dust distribution in our simulation.  
The synthetic images are shown in the upper panels of Figure \ref{fig:ALMA}. The left panel is the intensity
before convolving with the radio beam. The middle panel is after convolution with  the 
ALMA beam size of 0.32"$\times$0.21", which can be directly compared with Figure 1 of van der Marel \etal (2013) \footnote{Considering
the orientation of the ALMA observational beam is not horizontal as we assumed, we have rotated
the image slightly so that the relative position between the vortex and the observational beam is the same between simulations and observations.}. 
The only difference is that the peak intensity is half of ALMA observations by van der Marel \etal (2013).
This can be alleviated by assuming a slightly higher temperature or different dust size distribution. In the right panel, the beam size is
 0.16"$\times$0.11", similar to the beam size in ALMA cycle 2 observations. Thus, the vortex may have not been resolved in current
 ALMA observations and we predict that the vortex will look to be smaller
 in ALMA cycle 2 observations. 
 
We have used the same approach to synthesize ALMA observations for HD 142527.
We assume that the dust in HD 142527 has a power law distribution n(s)$\sim s^{-3.5}$ from 0.005 $\mu$m
to 150 $\mu$m. The total dust-to-gas mass ratio is assumed to be 1:50 to match the peak intensity in observations. 
Particles in size bins of [0.005$\mu$m, 4.4$\mu$m], [4.4$\mu$m, 14$\mu$m], [14$\mu$m, 44$\mu$m], 
and  [44$\mu$m, 100$\mu$m]  are represented by the surface density distributions of gas, Par. a, Par. b, and Par. c in our simulations. 
The opacities in these size bins are   0.12, 0.13, 0.26, 11.7  cm$^{2}$ g$^{-1}$ at 375 GHz.
The gas surface density follows  $\Sigma_{g}=22 (r/{\rm AU})^{-1}$ g cm$^{-2}$ and the temperature
follows $T = 354 (r/{\rm AU})^{-1/2}$~K (Casassus \etal 2013). The disk inclination angle is 20$^{o}$. The distance to the object is 140 pc. 
 ALMA observations by  Casassus \etal (2013) have a beam size of 0.51"$\times$0.33". The synthetic image is generally similar
 to ALMA observations (panel a of Figure 1 in Casassus \etal 2013) except that
 the shape of the vortex is more elongated. Again, we make
 predictions for ALMA cycle 2 observations with a beam size of  0.16"$\times$0.11", which shows a narrower rim
and possibly some emission from the horseshoe region along the planet's orbit. The synthetic image suggests that ALMA cycle 2 observations
will start to resolve the vortex structure in HD 142527.

\subsection{Implications}
If the non-axisymmetric density feature revealed by recent ALMA observations
is due to the gap edge vortex, it may indicate that
the protoplanetary disk beyond 10s of AU is indeed dominated by AD, as suggested by disk
ionization calculations (Bai \etal 2011a, 2011b, Perez-Becker \& Chiang 2011a, 2011b). 
The vortex has a significant radial extent ($\Delta r\sim$ 1 at $r=2$ for Par. b in Figure \ref{fig:imagegaspar}, compared with $h\sim 0.24$ at $r=2$), and
the presence of the vortex also makes the gap edge appear eccentric.
The vortex can exist over 1000 orbits, which is 2.5$\times$10$^{5}$ years
at 40 AU. This timescale is comparable to the lifetime of protoplanetary disks, and
thus consistent with observations.
With a different planet mass, $h/r$, or some other conditions, the vortex may be
able to survive for even longer (Fu \etal 2014). 

{  On the other hand, the vortex cannot last forever, and in the case we explored in Figure \ref{fig:imagegaslong}
 it disappears at 2000 orbits. }
Thus, it is not necessary that every transitional disk should have a gap edge vortex.
The planet induced gap edge could also be smooth in turbulent disks dominated by AD.
The presence of the gap edge vortex may indicate that AD dominates in the outer disk,
but the reverse is not necessary, that is the absence of the vortex does not suggest that AD is not dominating the outer disk dynamics.

{ Another implication of our results is that the ``C'' shaped horseshoe region
could also be observable.} Figure \ref{fig:ALMA} demonstrates that ALMA is capable of observing
the material within the horseshoe region. 
The upper left panel in Figure \ref{fig:imagegaslong} shows 
that the horseshoe region even has a noticeable
density enhancement in the gap at 1000 orbits. 

{ Besides these asymmetric features, we have also demonstrated that particles can be trapped 
at the planet induced gap edge in turbulent disks driven by the MRI.} Although this particle trapping by the gap edge
 has been suggested previously by both analytical
calculations (Rice \etal 2006; Pillina \etal 2012) and viscous hydrodynamical simulations (
Paardekooper \& Mellema 2004, 2006; Paardekooper 2007; 
Fouchet \etal 2007, 2010; Ayliffe \etal 2012; Zhu \etal 2012),
we demonstrate for the first time that it can occur in realistic protoplanetary disks with turbulence driven by the MRI. 

{  We also confirm in our MHD simulations that 
the gap can be wider for particles having $T_{s}$ closer to 1.} This suggests that observations at different wavelengths (e.g., 
near-IR, mid-IR, submm, mm, or cm) may find gaps having different radial extents. 
 In the outer part of a protoplanetary disk (e.g., $\sim$ 50 AU) where
$\Sigma_{g}\sim$ 1 or 10 g cm$^{-2}$, 1 cm particles have
$T_{s}\sim$1 or 0.1 from Equation (\ref{eq:ts}). 
Thus,   submm and cm observations may suggest wider gaps than 
observations at other wavelengths (e.g., near-IR) suggest.

{  Finally, we want to caution that there has not been any firm observational evidence 
that planets are present in transitional disks
although candidates exist (Kraus and Ireland 2012, Huelamo et al.
2011, and Cieza et al. 2013).
Photoevaporation is another proposed mechanism to open gaps and holes in transitional disks (see the review by Alexander et al. 2013).}
If  gas can pile up at the gap edge in the photoevaporation scenario, the gap edge is also subject to the RWI and leads to the formation of a vortex. 
However, unlike gap opening by planets (where the planet pushes disk material to the gap edge), 
the photoevaporation wind removes disk mass directly. Building a density bump or vortensity minimum may require
mass to return to the disk, or some thermodynamical processes (e,g, ``thermal sweeping'', Owen \etal 2013; ``photoelectric instability'', 
Lyra \& Kuchner 2013) that can lead to vortex formation.

\subsection{Caveats}
 
{ The main caveat in this study is the neglect of disk stratification, the Hall effect, and dust feedback.}
When the disk is threaded by net vertical fields, 
local stratified simulations suggest that the stratification does not affect the
 turbulent level at the midplane of the AD dominated disk (Simon \etal 2013). However, strong disk winds are developed 
(Bai 2013, Simon \etal 2013). It remains to be explored if these winds still develop in global stratified disks and can affect vortex formation.   
In reality, the disk could also have layered structure. If the active layer can transport
material fast enough, it may also prevent vortex generation (Lin 2014). Recently global MHD layered disk
simulations (Gressel \etal 2013) to study gap opening at 5 AU do not report the appearance of vortices at gap edges.
The Hall effect is small at 100 AU, but it could be important at 10 AU (Lesur \etal 2014, Bai 2014). 
Since the Hall effect can only be studied properly in stratified disks (Kunz \& Lesur 2013), 
carrying out global stratified simulations
with the Hall effect is needed to study the gap edge vortex at 10 AU.  

Our AD1e3 simulation suggests
that, at the vortex center, the surface density of dust Par. c ($t_{s}\Omega\sim 0.6$) is 0.1 times the gas surface density assuming 
the dust-to-gas mass ratio is 1:100.
Using $t_{eddy}\Omega \sim 3$ and $Sc_{z}\sim 0.3$ for this run (Zhu \etal 2014b),
the scale height of Par. c is 0.05 of the gas scale height. Thus, at the disk midplane,
the density ratio between dust and gas can reach to 2 if all dust has this size. Both 
2-D and 3-D unstratified simulations suggest that dust feedback can destroy
the vortex when the dust-to-gas mass ratio is large 
(Johansen \etal 2004). On the other hand, it is unclear how a large amount of dust at the
disk midplane can affect the dynamics of the vortex which extends through the whole disk from the midplane to the atmosphere.  
Future 3-D stratified simulations with dust feedback are desired. 
Although the gas disk is gravitationally stable (Toomre Q for the gas at 60 AU in Oph IRS 48 is $\sim$ 4000), the 
gas/dust two fluid system could be subject
to gravitational instability (Jog \& Solomon 1984), which requires more complicated simulations including
self-gravity of dust particles. 

{ The generation of a vortex at the gap edge depends on both planet and disk properties, and a
thorough parameter study is desired in future.}  The operation of the RWI requires the disk to have a vortensity minimum, or in other words,
a density bump at the gap edge. The planet pushes material to the gap edge leading to a density bump, while the viscosity or turbulent diffusion
tries to smooth out the density bump. The balance between these two processes determines whether the gap edge
is unstable to the RWI. The history of the planet growth is also important.
In our simulations, the planet is inserted in 10 orbits, while, in reality, the planet grows much slower. It is unclear if the vortex
can still be generated if the planet growth timescale is longer than the viscous timescale. 
Thus, carrying out a parameter study on both planet and disk properties is important to understand
the vortex formation at the gap edge in a realistic protoplanetary disk.

Finally we caution that the similarity between MHD and viscous simulations is limited to whether the vortex can be generated at the gap edge
under a given stress. This only implies that the density bump at the 
gap edge is smoothed by a strong turbulence or a large viscosity. The structure and
the longevity of the vortex can be quite different between MHD and viscous simulations, as shown in Figure \ref{fig:imagegaslong}.

\section{Conclusion}
We have performed a systematic study of particle trapping at the edge of a gap opened by a planet 
in a protoplanetary disk. In particular, we study the effects of turbulence driven by the magnetorotational instability (MRI) on particle trapping. We have carried out the first global
 3-D  magnetohydrodynamic (MHD) simulations including Lagrangian dust particles.
To study gap opening in the outer regions of protoplanetary disks, we also have included ambipolar diffusion in our MHD simulations.

Our work demonstrates that particles are trapped at the planet induced gap edge in MRI turbulent 
disks. The gap size is 
different for different particles, with the largest gap size for $T_{s}\sim 1$  particles. 

With ideal MHD,
strong turbulence 
can lead to non-axisymmetric density features at the planet induced gap edge for particles with $t_{s}\sim \Omega^{-1}$. However, the strong
turbulence in these cases also prevents vortex formation at the 
gap edge, and particles with $t_{s}\not\approx \Omega^{-1}$ are distributed in an axisymmetric fashion. 
When AD is considered, turbulence is significantly suppressed ($\alpha\lesssim 10^{-3}$), and the vortex
forms at the planet induced gap edge. The vortex can efficiently trap dust particles that span 3 orders of magnitude in size
within 100 planetary orbits, producing more than a factor of 10
enhancement in the surface density within the vortex. The existence of the vortex also makes the gap edge appear eccentric, with eccentricities up to 0.2. 
The vortex in our MHD simulations with AD lasts for $\sim$ 1000 orbits, which is 2.5$\times$10$^{5}$ years
at 40 AU, comparable to the protoplanetary disk life time.  We have also
carried out 2-D HD simulations using viscosity to approximate 
turbulence driven by the MRI. We find that 2-D
viscous HD simulations can reproduce vortex generation in  3-D MHD simulations very well. 

Finally, we use the  results from our first principle simulations to generate synthetic images for ALMA dust continuum observations 
on Oph IRS 48 and HD 142527, which show
good agreement with existing observations. The synthetic images also predict  
that the vortex structure will be smaller and narrower in higher resolution ALMA cycle 2 observations.
We conclude that 
the asymmetry in ALMA observations can be explained by dust trapping vortices and the existence of vortices
may be the evidence that the outer protoplanetary disks are dominated by ambipolar diffusion, as suggested by disk ionization calculations,  and the equivalent 
$\alpha$ is less than $10^{-3}$.

\acknowledgments
All simulations were carried out using
computers supported by the Princeton Institute of Computational Science and Engineering and Kraken at National 
Institute for Computational Sciences through XSEDE grant TG-AST130002. 
Z.Z. acknowledges support by
NASA through Hubble Fellowship grant HST-HF-51333.01-A
awarded by the Space Telescope Science Institute, which is
operated by the Association of Universities for Research in Astronomy, Inc., for NASA, under contract NAS 5-26555.
ZZ thanks Steve Lubow, Wen Fu, and Nienke van der Marel for helpful comments.

\end{document}